# Optical tristability and ultrafast Fano switching in nonlinear magneto-plasmonic nanoparticles

Wenjing Yu,[1,#] Pujuan Ma,[1,#] Hua Sun,[1,†] Lei Gao[1,2,*] and Roman E. Noskov[3,4]

[1]*College of Physics, Optoelectronics and Energy of Soochow University, Collaborative Innovation Center of Suzhou Nano Science and Technology, Soochow University, Suzhou 215006, China.*
[2]*Jiangsu Key Laboratory of Thin Films, Soochow University, Suzhou 215006, China.*
[3] *'Dynamics of Nanostructures' Laboratory, Department of Electrical Engineering, Tel Aviv University, Ramat Aviv, Tel Aviv 6139001, Israel.*
[4]*"Nanooptomechanics" Laboratory, ITMO University, St. Petersburg 197101, Russia.*
[#]*These authors contributed equally to this work.*
[†]*hsun@suda.edu.cn*
[*]*leigao@suda.edu.cn*

We consider light scattering by a coated magneto-plasmonic nanoparticle (MPNP) with a Kerr-type nonlinear plasmonic shell and a magneto-optic core. Such structure features two plasmon dipole modes, associated with electronic oscillations on the inner and outer surfaces of the shell. Driven in a nonlinear regime, each mode exhibits a bistable response. Bistability of an inner plasmon leads to switching between this state and a Fano resonance (Fano switching). Once the external light intensity exceeds the critical value, the bistability zones of both eigen modes overlap yielding optical tristability characterized by three stable steady states for a given wavelength and light intensity. We develop a dynamic theory of transitions between nonlinear steady states and estimate the characteristic switching time as short as 0.5 ps. We also show that the magneto-optical (MO) effect allows red- and blue- spectral shift of the Fano profile for right- and left- circular polarizations of the external light, rendering Fano switching sensitive to the light polarization. Specifically, one can reach Fano switching for the right circular polarization while cancelling it for the left circular polarization. Our results point to a novel class of ultrafast Fano switchers tunable by magnetic field for applications in nanophotonics.

## I. INTRODUCTION

Magneto-plasmonic architectures, consisting of plasmonic and MO elements, have recently become an active topic of research due to their multifunctionality stemming from magnetic tunability and nonreciprocity [1-7]. For instance, in magneto-plasmonic systems containing a metallic film perforated with subwavelength hole arrays and a dielectric film magnetized perpendicular to its plane, extraordinary transverse MO Faraday and Kerr effects were predicted [1]. Along this line, the giant enhancement of the transverse MO effect was experimentally found in magneto-plasmonic crystals composed of periodic gold nanowires structured on a thin layer of MO dielectric [8,9]. Furthermore, Belotelov et al. demonstrated MO-induced modulation of the transparency for magneto-plasmonic crystals [10]. Based on the Lorentz nonreciprocal model, Floess et al. derived the analytical expressions for a resonantly enhanced MO response in hybrid magneto-plasmonics and unraveled the underlying interplay between waveguide-plasmon-polariton quasiparticle and the Faraday rotation [5]. Beyond that, a great deal of attention has been payed to plasmonic nanorods in MO medium [6], magneto-plasmonic dimers [11,12], and magnetic nanoparticle arrays [13]. Additionally, a number of interesting effects has been predicted for core-shell MPNPs such as MO spasers [14], tunable plasmonic cloaks [15], enhanced Faraday rotation [16,17], circular dichroism [18] as well as the plasmon-driven Hall photon currents [19].

As is known, a coated nanoparticle encompassing a dielectric core and a concentric spherical plasmonic shell possesses two dipole plasmonic eigenmodes related to electronic oscillations localized on outer and inner shell surfaces [20]. Once they oscillate out-of-phase, the full particle dipole moment almost vanishes resulting in a Fano spectral shape of the scattering cross-section (because of that this mode also referred to as dark) [21]. In general, the Fano resonance (FR) features a steep, asymmetric line shape and an inherently excellent sensitivity to the changes in geometry compared to conventional resonance modes [22,23]. Due to large enhancement of the local fields, FR in a core-shell particle can boost the optical nonlinearity residing in a metallic shell and/or a dielectric core as it has been shown for a variety of structures demonstrating a Kerr-type nonlinear response [24-30]. The local field growth was also exploited to reach the strong control over the second-harmonic generation via the MO effect [31-33].

In this paper, we investigate the optical bistable/tristable behavior of coated MPNPs consisting of a MO core and a Kerr-type nonlinear plasmonic shell. In Section II we present the general theoretical model describing both the nonlinear steady-states and dynamical switching between them. Being in the framework of the quasistatic approximation, we account for the intrinsic nonlinear response of the metal via the self-consistent mean-field approximation [34-36] and find the steady-state solution. Then we employ the dispersion relation method [37-42] to develop the dynamical theory of switching. In Section III we show that bistability of plasmonic modes leads to tristability and switching between the bright high-energy plasmon mode and the dark Fano mode. We demonstrate the temporal dynamics of such transitions and estimate the characteristic switching time. Finally, we predict MO-induced tunability of the bistable spectral domains for the particle scattering cross-section. Specifically, red- and blue- spectral shifts of the Fano profile for right- and left- circular polarizations of the external light upon applying the external magnetic field are presented. Section IV summarizes the main results of the article.

## II. THEORETICAL MODELS

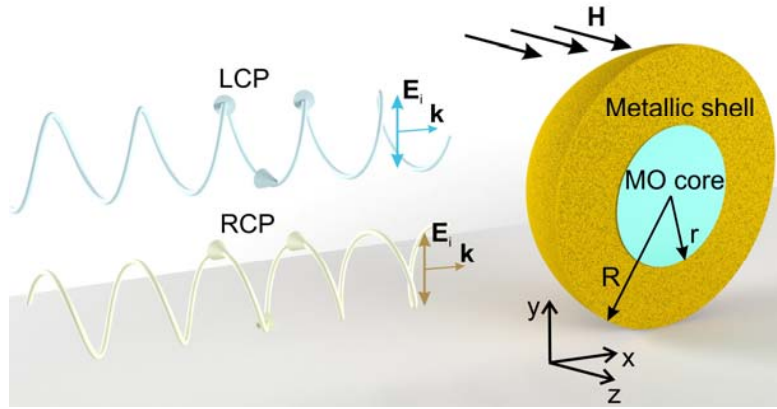

**FIG. 1.** (Color online) Schematics of the problem statement: left and right circularly polarized (LCP and RCP) light waves incident on a coated magneto-plasmonic nanoparticle (cross-section is only shown for clarity). The efficiency of light scattering can be switched between weak- and strong-scattering regimes by using a bistable response of plasmons localized on the inner and the outer shell surface. The static magnetic field **H** removes the degeneracy in scattering of LCP and RCP light, rendering nonlinear switching sensitive to the light polarization.

Figure 1 illustrates the considered core-shell nanoparticle: a MO (Bi:YIG) core with the permittivity tensor $\ddot{\varepsilon}_c$ and the radius $r$ coated by a silver shell (the radius $R > r$) with the nonlinear permittivity $\tilde{\varepsilon}_s = \varepsilon_s + \chi^{(3)} |\mathbf{E}_s|^2$, where $\varepsilon_s = \varepsilon_\infty - \omega_p^2/(\omega^2 + i\nu\omega)$ ($\varepsilon_\infty = 4.96$, $\hbar\omega_p = 9.54$ eV) is the linear permittivity, $\chi^{(3)} = 3 \times 10^{-9}$ esu is the typical cubic susceptibility of silver nanoparticles [43,44], and $|\mathbf{E}_s|^2$ the local intensity of the electric field inside the shell. The particle size is supposed to be much smaller than the light wavelength. Hereinafter we accept the harmonic time dependence $\exp(-i\omega t)$. We consider reduced losses $\hbar\nu = 0.0055$ eV (which can be achieved via decreasing the ambient temperature [45]) to get the pronounced Fano profile of the scattering cross-section. Similar particles have been extensively studied both experimentally [16,46] and theoretically [17-19,47]. We assume that the MO-core is magnetized by the external uniform magnetic field **H** along the $z$ axis. Thus, the gyrotropy of the core is described by a relative permittivity tensor,

$$\vec{\varepsilon}_c = \begin{pmatrix} \varepsilon & ig & 0 \\ -ig & \varepsilon & 0 \\ 0 & 0 & \varepsilon \end{pmatrix}.$$

Here the gyration $g = \chi_m H$ ($\chi_m$ is the magneto-optical susceptibility) is responsible for the "strength" of MO activity. We accept that the MO effect is induced by the external magnetic field only. For a typical MO-active material Bi:YIG $\varepsilon = 5.5 + 0.0025i$ [1] and the value of $g$ is tunable in the range of 0 - 0.3 by adjusting the magnetic field [48,49]. Since its 6 nm thick films demonstrate $\chi_m \approx 3.8 \times 10^{-5}$ esu [50], the maximal value of gyration which can be reached in practice corresponds to the magnetic field ~ 7.8 kG (or 0.78 T).

Note that here, for the sake of clarity, we ignore the metallic nonlocality which potentially may lead to spectral shifting of the Fano resonance peak and enhancement of the nonlinear response [51].

### Nonlinear steady-state solution

To begin with, we consider the linear problem (i.e. $\chi^{(3)} = 0$) within the quasistatic dipole approximation, supposing that the size of the nanoparticle is much smaller than the incident light wavelength. We look for the local electric field inside the core, the shell and outside the nanoparticle in the form [14,52]

$$\mathbf{E}_{core} = \hat{A}\mathbf{E}_i \tag{1a}$$

$$\mathbf{E}_{shell} = \hat{B}\mathbf{E}_i - \frac{\hat{C}\mathbf{E}_i}{\rho^3} + \frac{3(\hat{C}\mathbf{E}_i \cdot \mathbf{n})\mathbf{n}}{\rho^3} \tag{1b}$$

$$\mathbf{E}_{out} = -\frac{\hat{D}\mathbf{E}_i}{\rho^3} + \frac{3(\hat{D}\mathbf{E}_i \cdot \mathbf{n})\mathbf{n}}{\rho^3} + \mathbf{E}_i \tag{1c}$$

where $\mathbf{E}_i$ is the incident field, $\hat{A}$, $\hat{B}$, $\hat{C}$ and $\hat{D}$ are unknown antisymmetric tensors, and $\rho$ is the distance from the particle geometric center. Having applied the boundary conditions for Eqs. (1a)-(1c) on the inner and outer surfaces, we arrive at the following system of equations,

$$(\varepsilon\hat{I} + \hat{G})\hat{A} = \varepsilon_s \hat{B} + 2\varepsilon_s \frac{\hat{C}}{r^3}$$

$$\varepsilon_s \hat{B} + 2\varepsilon_s \frac{\hat{C}}{R^3} = 2\frac{\hat{D}}{R^3} + \hat{I}$$

$$\hat{A} = \hat{B} - \frac{\hat{C}}{r^3}$$

$$\hat{B} - \frac{\hat{C}}{R^3} = -\frac{\hat{D}}{R^3} + \hat{I}$$

where $\hat{G}$ is the nondiagonal part of the permittivity tensor $\vec{\varepsilon}_c$. These relations allow us to write the tensors $\hat{A}$, $\hat{B}$, $\hat{C}$ and $\hat{D}$ in the same form,

$$\hat{M} = \begin{pmatrix} M_{11} & iM_{12} & 0 \\ -iM_{12} & M_{11} & 0 \\ 0 & 0 & M_{33} \end{pmatrix} (M = A, B, C, D), \tag{2}$$

where

$$A_{11} = \frac{9R^3\varepsilon_s[2r^3(\varepsilon_s-1)(\varepsilon-\varepsilon_s)+R^3(\varepsilon_s+2)(\varepsilon+2\varepsilon_s)]}{P_+P_-},$$

$$A_{12} = \frac{-9gR^3\varepsilon_s[2r^3(\varepsilon_s-1)+R^3(\varepsilon_s+2)]}{P_+P_-},$$

$$B_{11} = \frac{3R^3\{g^2[2r^3(1-\varepsilon_s)-R^3(\varepsilon_s+2)]+(\varepsilon+2\varepsilon_s)[2r^3(\varepsilon-\varepsilon_s)(-1+\varepsilon_s)+R^3(\varepsilon_s+2)(\varepsilon+2\varepsilon_s)]\}}{P_+P_-},$$

$$B_{12} = \frac{18gR^3r^3\varepsilon_s(1-\varepsilon_s)}{P_+P_-},$$

$$C_{11} = \frac{3R^3r^3\{g^2[2r^3(1-\varepsilon_s)-R^3(\varepsilon_s+2)]+(\varepsilon-\varepsilon_s)[2r^3(\varepsilon-\varepsilon_s)(-1+\varepsilon_s)+R^3(\varepsilon_s+2)(\varepsilon+2\varepsilon_s)]\}}{P_+P_-},$$

$$C_{12} = \frac{9gR^6r^3\varepsilon_s(2+\varepsilon_s)}{P_+P_-},$$

$$D_{11} = \frac{R^3}{P_+P_-}\{[R^3(\varepsilon+2\varepsilon_s)(\varepsilon_s-1)+r^3(\varepsilon-\varepsilon_s)(1+2\varepsilon_s)][R^3(\varepsilon+2\varepsilon_s)(\varepsilon_s+2)+2r^3(\varepsilon-\varepsilon_s)(\varepsilon_s-1)]$$
$$-g^2[R^3(\varepsilon_s-1)+r^3(1+2\varepsilon_s)][R^3(\varepsilon_s+2)+2r^3(\varepsilon_s-1)]\},$$

$$D_{12} = \frac{27gR^6r^3\varepsilon_s^2}{P_+P_-}.$$

with $P_\pm = -2r^3(\varepsilon-\varepsilon_s)(1-\varepsilon_s)+R^3(2+\varepsilon_s)(\varepsilon+2\varepsilon_s)\pm g[2r^3(1-\varepsilon_s)-R^3(2+\varepsilon_s)]$. When $E_{iz}=0$, there exist two eigenvalues for the tensor $\hat{D}$, i.e.,

$$\alpha_\pm = D_{11}\pm iD_{12} = R^3\frac{r^3(2\varepsilon_s+1)(\varepsilon\mp g-\varepsilon_s)+R^3(-1+\varepsilon_s)(\varepsilon\mp g+2\varepsilon_s)}{2r^3(-1+\varepsilon_s)(\varepsilon\mp g-\varepsilon_s)+R^3(2+\varepsilon_s)(\varepsilon\mp g+2\varepsilon_s)} \quad (3)$$

where $-g$ corresponds to $\mathbf{E}_+ = (1\ i\ 0)^T$ polarization, while $+g$ to $\mathbf{E}_- = (1\ -i\ 0)^T$, and $\alpha_\pm$ is the particle polarizability corresponding to the left and right circularly polarized (LCP and RCP) light, respectively.

Next, we account for the Kerr-type nonlinearity of the metallic shell. In general, there is no an analytical solution for such problem as the local field in the shell is inhomogeneous, and one should solve numerically the nonlinear Laplace equation. However, when the nonlinear contribution is weak (i.e. $\varepsilon_s \gg \chi^{(3)}|\mathbf{E}_s|^2$), it can be considered in the first order of perturbation. This means that one can neglect nonlinearity-driven variance in the field structure and account for the self-action of the electromagnetic field via the averaged field [34]. Hence, we present the shell permittivity as

$$\tilde{\varepsilon}_s = \varepsilon_s + \chi^{(3)}|\mathbf{E}_s|^2 \approx \varepsilon_s + \chi^{(3)}\langle|\mathbf{E}_s|^2\rangle. \quad (4)$$

Here the field averaged in the shell volume $V_s$ can be expressed as

$$\langle|\mathbf{E}_s|^2\rangle = \frac{1}{V_s}\int_r^R\int_0^\pi\int_0^{2\pi}\mathbf{E}_s\mathbf{E}_s^*\rho^2\sin\theta d\rho d\theta d\phi \quad (5)$$

and the local electric field can be written in accord to Eqs. (1b) and (2) as,

$$\mathbf{E}_s = \begin{pmatrix} B_{11}E_{ix} + B_{12}E_{iy} \\ -B_{12}E_{ix} + B_{11}E_{iy} \\ B_{33}E_{iz} \end{pmatrix} - \frac{1}{\rho^3} \begin{pmatrix} C_{11}E_{ix} + C_{12}E_{iy} \\ -C_{12}E_{ix} + C_{11}E_{iy} \\ C_{33}E_{iz} \end{pmatrix}$$

$$+ \frac{3}{\rho^3} \begin{pmatrix} (C_{11}E_{ix} + C_{12}E_{iy})\sin\theta^2\cos\phi^2 + (-C_{12}E_{ix} + C_{11}E_{iy})\sin\theta^2\sin\phi\cos\phi + C_{33}E_{iz}\sin\theta\cos\theta\cos\phi \\ (C_{11}E_{ix} + C_{12}E_{iy})\sin\theta^2\sin\phi\cos\phi + (-C_{12}E_{ix} + C_{11}E_{iy})\sin\theta^2\sin\phi^2 + C_{33}E_{iz}\sin\theta\cos\theta\sin\phi \\ (C_{11}E_{ix} + C_{12}E_{iy})\sin\theta\cos\theta\cos\phi + (-C_{12}E_{ix} + C_{11}E_{iy})\sin\theta\cos\theta\sin\phi + C_{33}E_{iz}\cos\theta^2 \end{pmatrix},$$

where $\theta$ and $\phi$ are spherical polar and azimuthal angles.

When the incident light is LCP or RCP, i.e. $\mathbf{E}_i = E_0 e^{ikz}(1 \ \pm i \ 0)^T$, Eq. (5) can be simplified by using the spectral representation method [25,53] and written as follows,

$$\langle |\mathbf{E}_s|^2 \rangle = \frac{2E_0^2}{(2+\varepsilon \pm g)^2} \left( \frac{|X|^2}{|S-S_1|^2} + 2\eta^3 \frac{|Y|^2}{|S-S_2|^2} \right), \tag{6}$$

where $\eta = r/R$ is the aspect ratio between the core and the shell and

$$S = \frac{1}{1 - \varepsilon_s - \chi^{(3)} \langle |\mathbf{E}_s|^2 \rangle},$$

$$X = \frac{S[S(\varepsilon \pm g + 2) - 2]}{S - S_2},$$

$$Y = \frac{S[S(\varepsilon \pm g - 1) + 1]}{S - S_1}.$$

The poles $S_1$ and $S_2$ are given by,

$$S_1 = \frac{b - \sqrt{b^2 - 24(2+\varepsilon \pm g)(1-\eta^3)}}{6(2+\varepsilon \pm g)}, S_2 = \frac{b + \sqrt{b^2 - 24(2+\varepsilon \pm g)(1-\eta^3)}}{6(2+\varepsilon \pm g)}.$$

with $b = 8 + \varepsilon \pm g + 2\eta^3(\varepsilon \pm g - 1)$.

Now, to obtain the nonlinear particle polarizability one should exchange $\varepsilon_s$ with $\varepsilon_s + \chi^{(3)} \langle |\mathbf{E}_s|^2 \rangle$ in Eq. (3) so that Eqs. (3) and (6) form the self-consistent nonlinear steady-state solution.

### Dynamical model

To characterize the temporal behavior of our system, we derive the dynamical model for the MPNP response to the external electric field. To this end, we employ the dispersion relation method [37] adapted for plasmonic nanoparticles [38-41] and graphene flakes [42], and write the Fourier transforms of the MPNP electric-dipole moment as,

$$\alpha_\pm^{-1}(\omega)\mathbf{p}_\pm = \mathbf{E}_i, \tag{7}$$

where the subscript '$\pm$' corresponds to LCP and RCP. Assuming that $\chi^{(3)} \langle |\mathbf{E}_s|^2 \rangle \ll 1$ and $\nu/\omega_0 \ll 1$, we decompose $\alpha_\pm^{-1}(\omega)$ in the vicinity of the mode eigen frequencies $\omega_{1,2}$, and keep the first-order terms involving time derivatives to account for small broadening of the MPNP polarization spectrum,

$$\alpha_\pm^{-1} = \alpha_\pm^{-1}(\omega_{1,2}) + \frac{\partial \alpha_\pm^{-1}}{\partial \omega}\bigg|_{\omega=\omega_{1,2}} \left( \Delta\omega_{1,2} + i\frac{d}{dt} \right), \tag{8}$$

where $\Delta\omega_{1,2} = \omega - \omega_{1,2}$ is the frequency shift from the resonance value. The nonlinear term in Eq. (8) should be expressed via the particle dipole moment to reach the self-consistent model. Therefore, we use Eq. (6) at $\chi^{(3)} = 0$ and $E_0 = \alpha_\pm^{-1}(\omega_{1,2}) p_\pm$.

Next, we substitute Eq. (8) into Eq. (7) and obtain the governing dynamical equations for the slowly varying amplitudes of the particle polarization $P_{1,2}^\pm$ corresponding to the plasmon modes localized in the outer and inner surfaces of the silver shell,

$$i\frac{dP_1^\pm}{d\tau} + \left(i\gamma_1 + \Omega_1 + \left|P_1^\pm\right|^2\right)P_1^\pm = E,$$

$$i\frac{\omega_1}{\omega_2}\frac{dP_2^\pm}{d\tau} + \left(i\gamma_2 + \Omega_2 + \varsigma^2\left|P_2^\pm\right|^2\right)\varsigma P_2^\pm = \kappa E,$$

(9)

where $\tau = \omega_1 t$, $\Omega_{1,2} = \Delta\omega_{1,2}/\omega_{1,2}$, and $\gamma_{1,2} = \text{Im}\{\alpha_\pm^{-1}(\omega_{1,2})\}/\left(\partial_\omega \alpha_\pm^{-1}\big|_{\omega=\omega_{1,2}} \cdot \omega_{1,2}\right)$. Here we use the following normalization:

$$E = E_i \left(\psi_1 \xi_1\right)^{1/2} \left(\partial_\omega \alpha_\pm^{-1}\big|_{\omega=\omega_1} \cdot \omega_1\right)^{-3/2},$$

$$P_{1,2}^\pm = p_{\pm 1,\pm 2} \left(\psi_1 \xi_1 / \left(\partial_\omega \alpha_\pm^{-1}\big|_{\omega=\omega_1} \cdot \omega_1\right)\right)^{1/2},$$

$$\psi_{1,2} = \frac{\left[2r^3(\varepsilon \mp g + 1 - 2\varepsilon_s) + R^3(4 + \varepsilon \mp g + 4\varepsilon_s)\right]\chi^{(3)}}{\left[r^3(2\varepsilon_s + 1)(\varepsilon \mp g - \varepsilon_s) + R^3(\varepsilon \mp g + 2\varepsilon_s)(\varepsilon_s - 1)\right]R^3}\bigg|_{\substack{\omega=\omega_{1,2}\\ \text{Im}\{\varepsilon_s\}=0 \\ \text{Im}\{\varepsilon\}=0}},$$

$$\xi_{1,2} = \frac{18R^3\left(2r^3(\varepsilon \mp g - \varepsilon_s)^2 + R^3(\varepsilon \mp g + 2\varepsilon_s)^2\right)}{\left([r^3(2\varepsilon_s + 1)(\varepsilon \mp g - \varepsilon_s) + R^3(-1+\varepsilon_s)(\varepsilon \mp g + 2\varepsilon_s)]R^3\right)^2}\bigg|_{\substack{\omega=\omega_{1,2}\\ \text{Im}\{\varepsilon_s\}=0 \\ \text{Im}\{\varepsilon\}=0}}.$$

The terms

$$\kappa = \frac{\left(\psi_2 \xi_2\right)^{1/2} \left(\partial_\omega \alpha_\pm^{-1}\big|_{\omega=\omega_2} \cdot \omega_2\right)^{-3/2}}{\left(\psi_1 \xi_1\right)^{1/2} \left(\partial_\omega \alpha_\pm^{-1}\big|_{\omega=\omega_1} \cdot \omega_1\right)^{-3/2}},$$

$$\varsigma = \left(\frac{\psi_1 \xi_1}{\partial_\omega \alpha_\pm^{-1}\big|_{\omega=\omega_1} \cdot \omega_1}\right)^{-1/2} \left(\frac{\psi_2 \xi_2}{\partial_\omega \alpha_\pm^{-1}\big|_{\omega=\omega_2} \cdot \omega_2}\right)^{1/2}$$

characterize the system dispersion.

Finally, to get the full system response, one should summarize solutions of Eq. (9), i.e. $P^\pm = P_1^\pm + P_2^\pm$. For example, the steady state solution is given by

$$\left(i\gamma_1 + \Omega_1 + \left|P_1^\pm\right|^2\right)P_1^\pm = E,$$

$$\left(i\gamma_2 + \Omega_2 + \varsigma^2\left|P_2^\pm\right|^2\right)\varsigma P_2^\pm = \kappa E.$$

(10)

Being physically identical to the nonlinear solution presented by Eqs. (3) and (6), Eq. (10) describes the Fano resonance, bistability, multistability as well as MO-tuning of the nonlinear properties.

## III. RESULTS AND DISCUSSION

### Analysis of the nonlinear steady-state solution

To provide the quantitative characterization for the light-MPNP interaction, we introduce the light scattering efficiency which is defined as $Q_{sc} = 8/3(2\pi R/\lambda)^4 |\Delta_1|^2$, with $\Delta_1 = \alpha_\pm / R^3$ [54], where $\alpha_\pm$ can be given by Eqs. (3) and (6) or by $\alpha_\pm = (P^\pm / E)\left[\partial_\omega \alpha_\pm^{-1}\Big|_{\omega=\omega_1} \omega_1\right]^{-1}$.

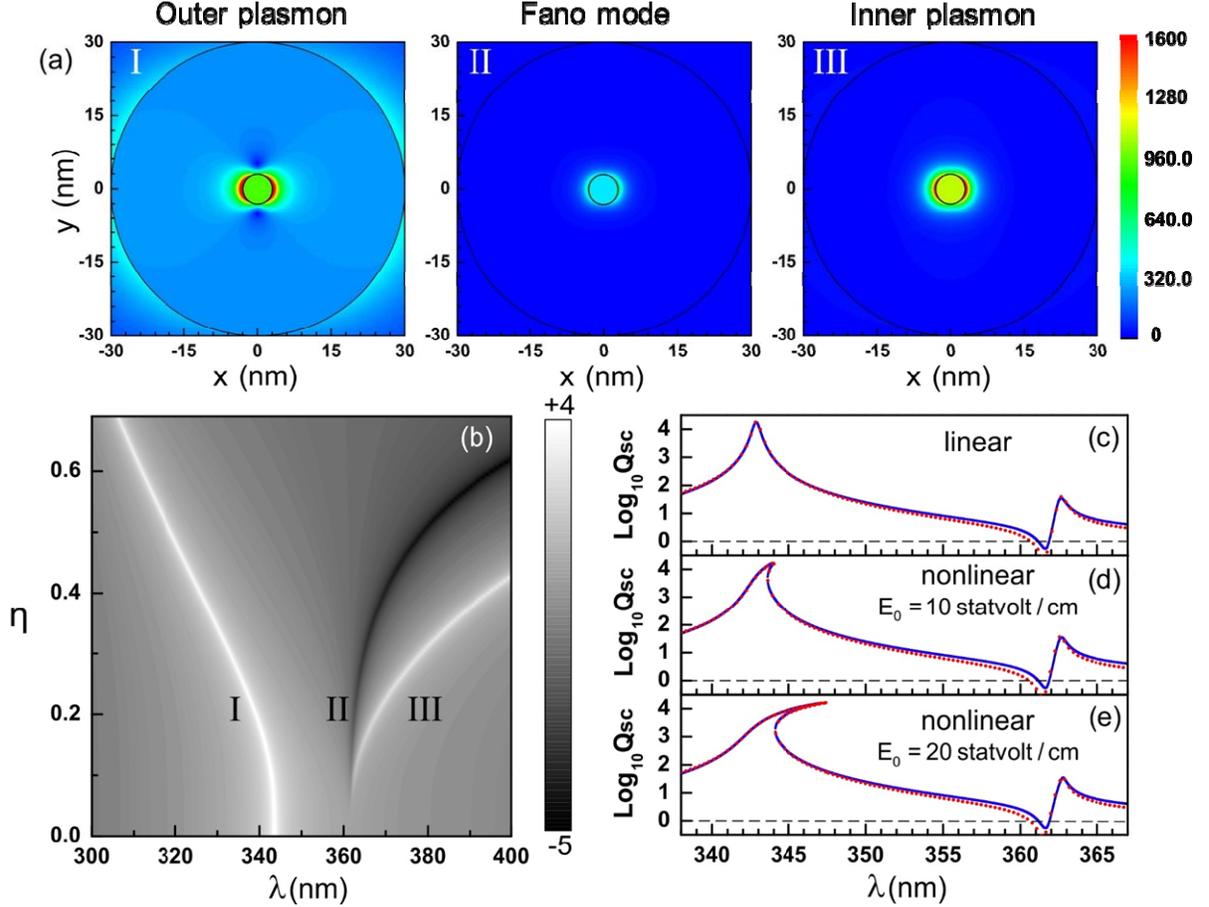

FIG. 2. (Color online) (a) The electric field structure for the eigen modes and the Fano mode of the core-shell particle when $\eta = 0.1$. Mode labels correspond to (b). (b) Linear light scattering efficiency as a function of wavelength and aspect ratio in the logarithmic scale ($\log_{10}(Q_{sc})$). (c), (d) and (e) Scattering efficiency spectra at different external fields for $\eta = 0.1$ and $R = 30$ nm. Continuous blue and dotted red curves correspond to solutions given by the direct averaging (Eqs. (3) and (6)) and the dispersion relation method (Eq. (10)).

First, we consider a linear nonmagnetic regime. Figure 2(a) plots the electric field structure for the different modes of the MPNP. There exist two bright dipole resonant modes, associated with electronic oscillations on the surfaces of the shell (outer and inner plasmons), and the dark Fano mode arisen as a result of out-of-phase oscillations of these modes. The relative spectral position of the mode resonant wavelengths depends on the ratio between permittivities of the core and the host medium [51]. Since in our case $\varepsilon > 4\varepsilon_h$ ($\varepsilon_h = 1$), the eigen wavelength of the sphere-like plasmon localized on the outer surface is smaller than the one of the cavity-like plasmon localized on the inner surface ($\lambda_s < \lambda_c$) [Figs. 2(b) and (c)]. The aspect ratio defines the strength of mode coupling evidenced by the eigen wavelengths splitting. For $\eta \leq 0.1$ (the thick shell), the plasmons are almost decoupled and the lines of the

Fano mode and the inner plasmon approach to each other yielding a narrowband Fano-shape curve [51]. As the shell gets thinner, the mode interaction growths, leading to strong splitting the mode wavelengths. It would be instructive to note that in case $\varepsilon < 4\varepsilon_h$, the Fano-shape profile appears, in sharp contrast to our system, in the vicinity of the outer plasmon [21].

Next, we investigate the impact of Kerr-type nonlinearity on the scattering efficiency still in a nonmagnetic regime. As the light intensity grows, nonlinearity firstly gives rise to bistability for the outer plasmon dipole mode along with almost zero effect on the Fano resonance [see Figs. 2(d) and (e)]. Importantly, such behavior can be reached for comparatively low optical fields $\sim 20$ statvolt/cm $(6\times 10^5 \text{ V/m})$ due to the synergy of the resonance local field enhancement and the high cubic metallic susceptibility.

It is methodologically useful to compare solutions given by the direct averaging and the dispersion relation method. They are shown in Figs. 2(c)-(e). Remarkably, since both these techniques work in the first-order perturbation theory, they yielded almost identical results.

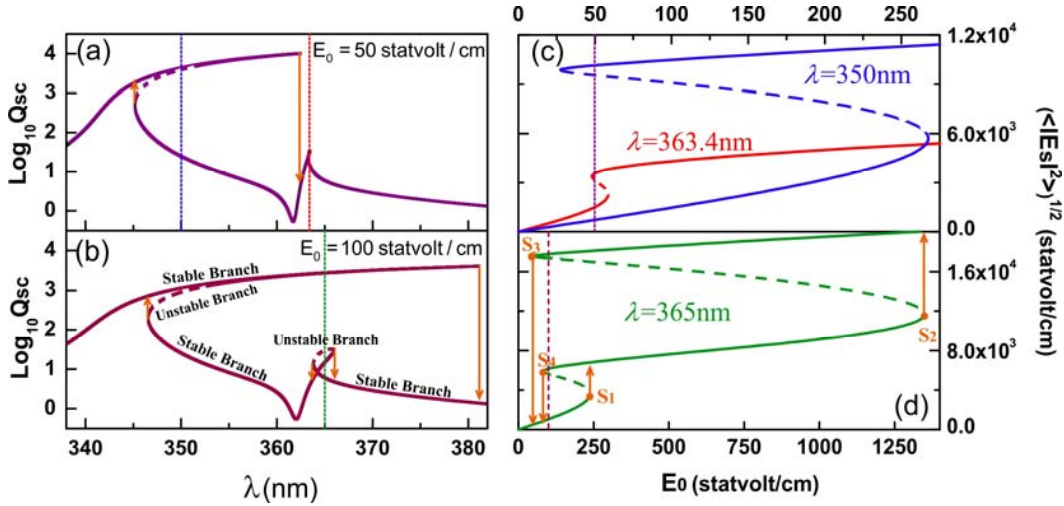

**FIG. 3.** (Color online) (a) and (b) Scattering efficiency spectra in the nonlinear regime for different values of the driving electric field. (c), (d) The average optical field in the shell as a function of the incident field for different wavelengths indicated in (a) and (b) by dashed lines. Vertical dashed lines indicate the fields for which (a) and (b) were plotted. All other parameters are the same as in Fig. 2.

Once $E_0$ exceeds the critical value $\sim 50$ statvolt/cm $(1.5\times 10^6 \text{ V/m})$, both dipole modes demonstrate optical bistable regions (OBRs) [Fig. 3(a)]. Additionally, one can undergo a transition between the bright high-energy dipole mode and the dark Fano mode once the bistability domain is wide enough. For stronger fields, OBRs get overlapped yielding the optical tristable region (OTR) or optical tristability characterized by 3 stable steady states [see Fig. 3(b)]. These states correspond to resonant dipole plasmons and an off-resonance response. Appearance of two separate OBRs and OTR can also be observed in Figs. 3(c) and 3(d), which show the average field in the shell vs the driving field $E_0$ for different wavelengths.

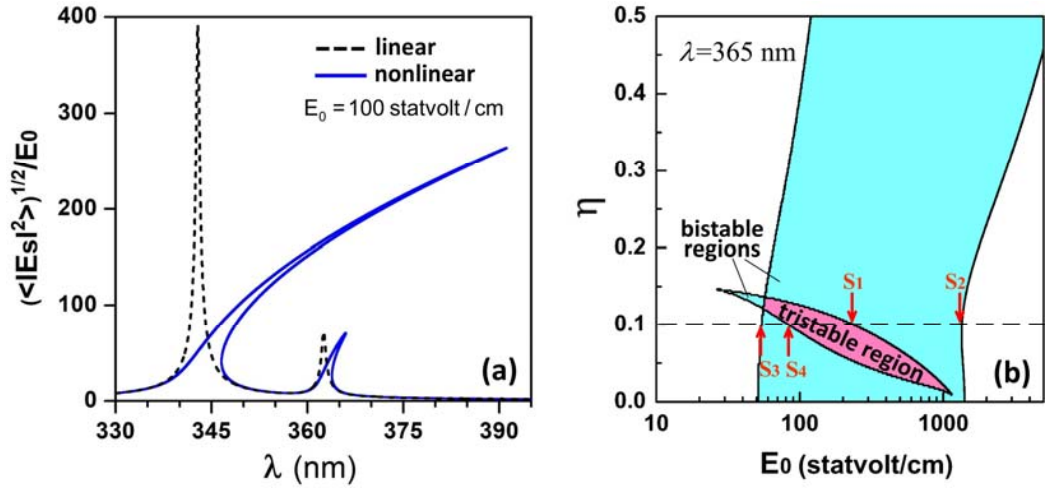

**FIG. 4.** (Color online) (a) The enhancement factor $\langle |E_s|^2 \rangle^{1/2} / E_0$ vs the wavelength in the linear and non-linear regimes. (b) Bifurcation diagram showing bistability and tristability zones as functions of volume fraction $\eta$ and the incident field $E_0$ for $\lambda = 365$ nm. The points $S_1$-$S_4$ correspond to the bistability thresholds marked in Fig. 3(d).

Importantly, the resonant enhancement of the local field inside the shell may reach the factor of several hundreds relative to the incident field, which is shown in Fig. 4(a). This allows us to reach OBRs and OTR at the moderate optical field facilitating the usage of the proposed core-shell nanoparticles as nonlinear nanoswitchers. We also point out that OTR can be obtained only for the wavelength larger than the eigen wavelength of the inner plasmon mode. This stems from the fact that the Kerr-type metallic nonlinearity has a focusing type resulting in the red-shift of the eigen wavelengths as the optical intensity grows, as shown in Fig. 4(a).

Let us now analyze the impact of the shell thickness on bistability and multistability. To this end, we plot the bifurcation diagram showing bistable and tristable regions for a fixed wavelength near the Fano resonance, as a function of the volume fraction $\eta$ and the incident field $E_0$ in Fig. 4(b). When the shell is significantly thicker than the core ($0 < \eta \leq 0.137$), the plasmons are weakly interacting and their eigenfrequencies are not shifting as $\eta$ growths [Fig. 2(b)]. In this situation, one can observe clear Fano resonance and overlapping OBRs of both dipole modes leading to tristability. For example, at $\eta = 0.1$ [dashed line in Fig. 4(b)] the threshold fields are consistent with the switching points $S_1$-$S_4$ marked in Fig. 3(d). However, decreasing the shell thickness gives rise to the strong plasmon hybridization, which shifts eigenfrequencies from each other [Fig. 2(b)]. As a result, one cannot reach OTR at the realistic optical power.

Next, we analyze tuning the scattering efficiency by the gyrotropy of the core. The MO effect exerts pronounced shifts in the eigen wavelength of the inner plasmon; while having almost no impact on its counterpart [Fig. 5(a)]. Hence, the associated Fano line experiences blue- and red-shift for LCP and RCP waves, respectively.

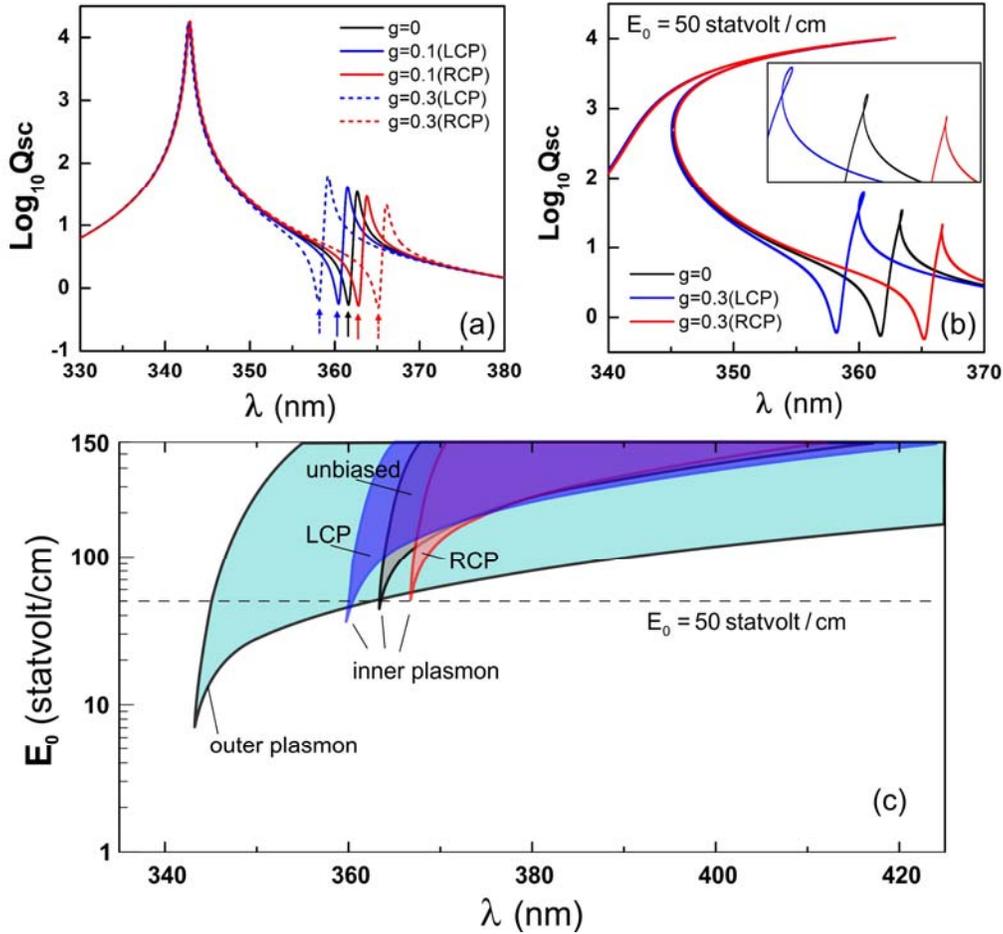

**FIG. 5.** (Color online) (a) Linear and (b) nonlinear scattering efficiency spectra for right- and left-circular polarized light in presence of MO effect. Inset shows the resonance lines for the inner plasmon in detail. (c) Bistability/tristability zones in $(E_0, \lambda)$-axes for non-magnetic and magnetic ($g = 0.3$) regimes. Magnetization just slightly shifts the bistability zone for the outer plasmon (unnoticeable in a logarithmic scale). For all figures $\eta = 0.1$.

Figures 5 (b) and (c) illustrate the impact of the gyrotropy on the optical bistability and multistability. In absence of $\mathbf{H}$ ($g = 0$), LCP and RCP light yield equal scattering efficiency [the same as in Fig. 3(a)]. Although the MO effect almost does not influence the bistability zone of the outer plasmon, the Fano line and the bistability domain of the inner plasmon get considerably blue- and red-shifted for the LCP and RCP wave. This, in particular, yields an interesting opportunity of reaching tristability for LCP while cancelling it for RCP for a particular band of wavelengths [Fig. 5(b)]. In addition, the bistability field threshold for LCP drops but for RCP it rises [Fig. 5(c)]. Thus, gyrotropy removes the degeneracy in a nonlinear optical response of MPNPs for LCP and RCP driving, which can be used in polarization-sensitive switching discussed in the next Section.

**Temporal dynamics**

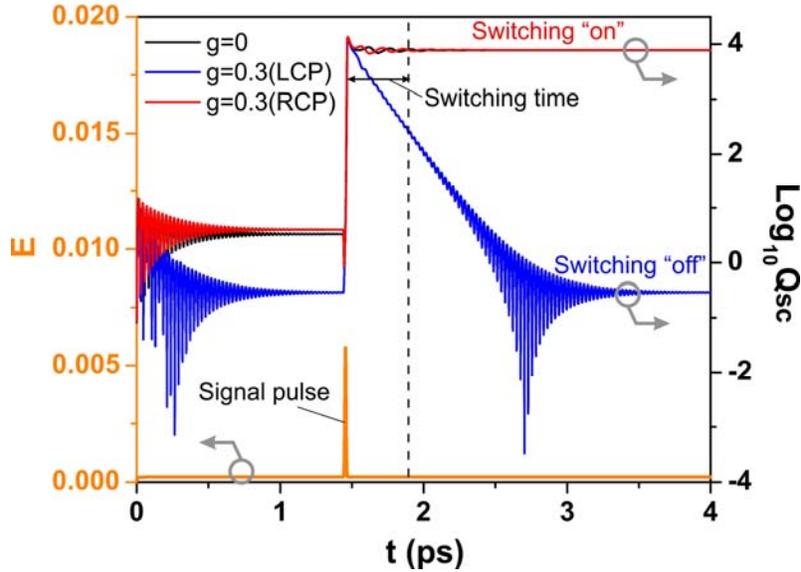

**FIG. 6.** (Color online) Temporal dependencies of the scattering efficiency and the dimensionless driving field (orange). The signal pulse provokes a transition from a weak-scattering to a strong-scattering state for $g=0$ (black) and $g=0.3$ in case of a right circularly polarized light (red). For a left circularly polarized light the system returns to the initial state (blue). The parameters are $E_0 = 50$ statvolt/cm (the background field), $E_p = 1265$ statvolt/cm (the pulse peak field) and $\lambda = 358$ nm.

In this section we study the temporal nonlinear dynamics of the MPNP described by Eq. (9). In Fig. 6, we plot the temporally dependent scattering efficiency for both LCP and RCP waves with a fixed wavelength close to the Fano resonance [see Fig. 5(b)]. The background field starts to growth from zero at $t=0$ reaching the saturation level $E_0 = 50$ statvolt/cm. In a while (~ 1 ps) the system comes into the steady state. Then a Gaussian signal pulse centered at 1.46 ps appears. In a nonmagnetic regime and for $g=0.3$ in case of RCP driving, the system transits from a weak-scattering to a strong-scattering steady state with a characteristic switching time ~ 0.5 ps. Remarkably, similar characteristic switching times were obtained for semiconductor microcavities [55,56]. However, for LCP the switching is cancelled and the MPNP comes back to the initial steady state. This happens because for a chosen wavelength 358 nm the peak signal pulse field overcomes the bistability thresholds for $g=0$ and $g=0.3$ in case of RCP but does not large enough to induce a transition for LCP. Although for the outer plasmon, the steady state bistability domain is just slightly affected by gyrotropy [Figs. 5(b) and (c)], this effect occurs to be pronounced enough to induce difference in dynamical switching.

Finally, we estimate the maximal driving time to avoid the particle ablation. To this end, we rely on the results of the previous study on the ablation thresholds for silver particles providing value about 3.96 J/cm$^2$ [57]. Taking into account the amplification of the electric field inside the Ag shell due to plasmonic resonance, we come to the critical illumination duration after which the MPNP is expected to be burned ~ 330 ps, which is much longer than the characteristic switching time. Thus, the all predicted phenomena are readily to be observed in experiment.

## IV. CONCLUSION AND OUTLOOK

In summary, we have shown that a coated magneto-plasmonic nanosphere with a Kerr-type nonlinear plasmonic shell and a magneto-optic core features optical bistable responses for dipole-type plasmons localized on the outer and inner surfaces of the shell. This bistability can be used to induce switching between the bright outer plasmon and a dark Fano resonance (Fano switching). Once the intensity of the driving light exceeds the critical value, optical tristability arises as a result of overlapping the bistability zones of both eigen modes. With help of the dispersion relation method, we have developed a dynamic theory of transitions between nonlinear steady states and estimated the characteristic switching time as short as 0.5 ps. The MO effect allowed spectral red- and blue-tuning of the Fano profile for right and left circularly polarized light, making Fano switching possible for RCP and cancelling it for LCP.

Our results pave the way for using MO elements for tuning nonlinear dynamical response of nanostructures. For example, nonlinear magnetoactive switchers sensitive to the light polarization can significantly increase the number of simultaneously processing channels for on-chip nanoantennas [58]. In analogy to metallic nanoparticles, arrays of MPNPs are expected to support a variety of nonlinear subwavelength dynamical modes in the form of kinks, solitons, oscillons and spatial patterns [40,59-62], and the MO effect will serve the powerful tool for steering these modes and searching for novel nonlinear dynamical phenomena.

**ACKNOWLEDGMENTS**

This work was supported by the National Natural Science Foundation of China (Grant No. 11374223), the National Science of Jiangsu Province (Grant No. BK20161210), the Qing Lan project, "333" project (Grant No. BRA2015353), and PAPD of Jiangsu Higher Education Institutions. The calculations of temporal dynamics for the particle scattering efficiency have been supported by the Russian Science Foundation Grant No. 16-12-10287. R.E.N. acknowledges support from Russian Foundation for Basic Research (Grant No. 16-02-00547) and hospitality of Prof. L. Gao and Soochow University.